\documentclass[12pt]{iopart}

\usepackage{iopams}
\usepackage{graphicx}

\usepackage{xcolor}
\expandafter\let\csname equation*\endcsname\relax
\expandafter\let\csname endequation*\endcsname\relax
\usepackage{amsmath}
\usepackage{bm}

\begin{document}

\title{Mean field theory of jamming of nonspherical particles}

\author{ Harukuni Ikeda$^{1,2}$, Pierfrancesco Urbani$^3$, and Francesco
Zamponi$^{1,2}$}

\address{
$^1$Laboratoire de Physique de l'Ecole Normale Sup\'erieure, CNRS, Paris,
France \\
$^2$ENS, Universit\'e PSL, Sorbonne Universit\'e, Universit\'e Paris-Diderot,
Sorbonne Paris Cit\'e, Paris, France \\
$^3$ Institut de physique th\'eorique, Universit\'e Paris Saclay, CNRS, CEA, F-91191 Gif-sur-Yvette, France
}
\ead{harukuni.ikeda@ens.fr}
\vspace{10pt}
\begin{indented}
\item[]March 2019
\end{indented}

\begin{abstract}
Recent computer simulations have uncovered the striking difference
between the jamming transition of spherical and non-spherical
particles. While systems of spherical particles are isostatic at the
jamming point, systems of nonspherical particles are not: the contact
number and shear modulus of the former exhibit a square root singularity
near jamming, while those of the latter are linearly proportional to the
distance from jamming. Furthermore, while our theoretical understanding
of jamming of spherical particles is well developed, the same is not
true for nonspherical particles.  To understand jamming of non-spherical
particles, in the previous work [Brito, PNAS 115(46), 111736], we
extended the perceptron model, whose SAT/UNSAT transition belongs to the
same universality class of jamming of spherical particles, to include
additional variables accounting for the rotational degrees of freedom of
nonspherical particles. In this paper, we give more detailed
investigations of the full scaling behavior of the model near the
jamming transition point in both convex and non-convex phases.

\end{abstract}

\newcommand{\diff}[2]{\frac{d#1}{d#2}}
\newcommand{\pdiff}[2]{\frac{\partial #1}{\partial #2}}
\newcommand{\fdiff}[2]{\frac{\delta #1}{\delta #2}}
\newcommand{\new}{\nonumber\\}
\newcommand{\bx}{\bm{x}}
\newcommand{\bu}{\bm{u}}
\newcommand{\hr}{\hat{r}}
\newcommand{\hf}{\hat{f}}
\newcommand{\hpi}{\hat{\pi}}
\newcommand{\bX}{\bm{X}}
\newcommand{\de}{\mathrm{d}}

\newcommand{\tk}{\tilde{k}}
\newcommand{\abs}[1]{\left|#1\right|}
\newcommand{\ave}[1]{\left\langle #1\right\rangle}
\newcommand{\M}{\mathcal{M}}
\newcommand{\Y}{\mathcal{Y}}
\newcommand{\A}{\mathcal{A}}
\newcommand{\Z}{\mathcal{Z}}
\newcommand{\G}{\mathcal{G}}
\newcommand{\PP}{\mathcal{P}}
\newcommand{\he}{\mathcal{H}}
\newcommand{\im}{{\rm Im}}
\newcommand{\erf}{{\rm erf}}

\newpage

\section{Introduction}

A system consisting of macroscopic particles that are so large that the
thermal fluctuations are negligible, at small packing fraction $\varphi$
easily deforms and flows under the action of external driving forces,
such as shear and shake. Upon increasing $\varphi$, the constituent
particles begin to be in contact and the system has a finite rigidity at
a certain critical packing fraction $\varphi=\varphi_J$. This
non-equilibrium fluid-solid transition is the so-called jamming
transition and $\varphi_J$ is referred to as the jamming
point~\cite{van2009,torquato2010}.  The jamming transition has been the
focus of an intense research activity, as it is ubiquitously observed
in a wide range of engineering and biological systems such as metallic
balls~\cite{bernal1960}, foams~\cite{durian1995foam,ohern2003},
colloids~\cite{zhang2009}, polymers~\cite{karayiannis2009},
candies~\cite{donev2004}, dices~\cite{jaoshvili2010} and
tissues~\cite{bi2015}.

One of the most popular and simple models to study the jamming
transition is a system consisting of frictionless spherical particles
with purely repulsive interaction. A large number of numerical works
have been performed on this model which have proven that (i) this system
is isostatic at $\varphi_J$, meaning that the contact number between
constituent particles is identical to the number the degrees of freedom,
so that the system barely achieves mechanical
stability~\cite{bernal1960}, (ii) the distributions of inter particle
distances and forces exhibit power law behaviors at
$\varphi_J$~\cite{donev2005, lerner2013low, wyart2012marginal}, (iii)
various physical quantities, such as the shear modulus and contact
number, exhibit power law scaling near $\varphi_J$~\cite{ohern2003}, and
(iv) a set of lengthscales diverging at $\varphi_J$~\cite{silbert2005, HLN18, HUZ19}.
These results suggest that the jamming transition point can be
considered as a sort of out-of-equilibrium critical point.

The theoretical understanding of jamming of spherical particles is also
quite advanced.  On the one hand, M.~Wyart \textit{et al.}  developed
a variational argument showing that isostaticity plays a crucial
role in determining the critical properties of the jamming
transition~\cite{WNW05,Wy12}.  On the other hand, a system consisting of
spherical particles has been solved exactly in the large dimension limit
by using the replica liquid theory (RLT)~\cite{MP09,parisi2010}. The RLT
is a theory that combines the density functional theory of classical
amorphous solids~\cite{SSW85,KW87,KT89} with the replica
method~\cite{mezard1987}, which is a standard method to treat disordered
systems originally developed to study spin glasses.  The RLT predicts
that a system near $\varphi_J$ is located in the (full) replica
symmetric breaking (RSB) phase~\cite{kurchan2013exact}. In the RSB
phase, the system is marginally stable and has a gapless spectrum of
excitations~\cite{franz2015,altieri2016jamming}. Interestingly, this
marginal stability is directly related to the isostaticity of the
jamming transition point~\cite{charbonneau2014fractal}. The exact
solution of hard spheres provides the exact values of the critical
exponents of the gap and force distribution function at $\varphi_J$ in
infinite dimensions, whose values agree well with the numerical results
in two and three dimensions within the numerical
precision~\cite{charbonneau2014fractal,charbonneau2014exact,charbonneau2015jamming,charbonneau2017glass}.

Furthermore in recent years, Franz and Parisi have proposed to look at
the jamming transition point as the satisfiability transition point for
continuous constraint satisfaction problems \cite{franz2016simplest}. In
a constraint satisfaction problem one has to find a configuration of a
set of dynamical variables that satisfy a set of constraints.  As the
number of constraints increases, one can observe a sharp phase
transition from a satisfiable (SAT) phase, where there are
configurations of the variables satisfying all the constraints, to an
unsatisfiable (UNSAT) phase, where no such configurations exists and a
finite fraction of constraints are
violated~\cite{monasson1999,mezard2002analytic,krzakala2007gibbs}. The
jamming transition of spherical particles can be considered as a
SAT-UNSAT transition within a glass: in the unjammed phase, one can find
disordered configurations satisfying the constraints that spheres do not
overlap, meaning that the system is in the SAT phase, while in the
jammed phase, one can not find such configurations and the number of
contacts between particles is finite, meaning that the system is in the
UNSAT phase~\footnote{Note that the jamming and glass transitions are
different phenomena: the glass transition is defined as the point at
which the system lose the ergodicity, while the jamming transition is
the point at which the particle can not avoid overlap.  }. In
\cite{franz2016simplest} Franz and Parisi considered the simplest
continuous constraint satisfaction problem, the perceptron, and showed
that the gap and force distributions close to the non-convex SAT/UNSAT
(jamming) threshold have the same critical exponents of those of hard
spheres in the large dimension limit, implying that the two models
belong to the same universality class. The simplicity of the perceptron
also allows one to calculate the contact number and the density of
states in the jammed (UNSAT) phase, and successfully reproduce the
numerical results on jamming of spheres in finite dimensions (even if
the origin of this universality is still unclear).

However, spherical particles are an idealized model. In general,
asphericity is inevitably introduced in realistic situations (where also
friction plays a role, although it will be always neglected here).  It
has been shown in the last years that the particle shape affects the
properties of jamming.  Numerical simulations on jamming of ellipsoids
shows that the contact number $z_J$ at jamming smoothly increases as the
asphericity increases, meaning that the system is not
isostatic~\cite{donev2004,mailman2009}. The breakdown of isostaticity is
also observed for spherocylinders~\cite{williams2003},
superballs~\cite{jiao2010}, superellipsoids~\cite{delaney2010}, other
convex shaped particles~\cite{werf2018} and even deformable
polygons~\cite{boromand2018}. Furthermore, detailed numerical
simulations showed that the critical exponents of the contact number and
shear modulus near the jamming transition point differ quite
significantly from those of spherical
particles~\cite{mailman2009,schreck2012}.  This difference suggests that
the universality class of the jamming transition of nonspherical
particles is different from the one of spherical objects.

In recent years, a series of phenomenological and theoretical approaches have emerged
to rationalize jamming of non-spherical particles.
For instance, by analyzing the
stability matrix, Donev \textit{et al.}~\cite{donev2007} argue that the breakdown of the
isostaticity is caused by the positiveness of the pre-stress
term.  By using the Edwards ensemble approach, Baule
\textit{et al.}~\cite{baule2013mean}  calculated the contact number and the jamming
transition point of nonspherical particles, which can qualitatively
reproduce the numerical results. However, to the
best of our knowledge there is no unified theoretical understanding that
explains the origin of the different critical exponents and universality
for nonspherical particles.
 In this work, we study a model that we have
recently proposed in a shorter report~\cite{brito2018universality} as a
simple exactly solvable mean-field model for the jamming transition of
nonspherical particles and we detail its analytical solution. We
consider the perceptron model, introduced in \cite{franz2016simplest} to
study the jamming transition of spherical particles, and extend it to
include additional dynamical variables, playing the same role as the
rotational degrees of freedom of nonspherical particles
\cite{brito2018universality}. The model and its scaling properties can
be investigated analytically by using the replica method, as in the case
of the original perceptron model~\cite{franz2017}. We find that the gap
and force distributions do not exhibit power law behavior and remain
regular even at the jamming transition point. This is in marked
constrast with the original perceptron model or spherical particles,
where the gap and force distributions exhibit power
laws~\cite{franz2017}. We find that the absence of criticality of the
gap and force distributions are originated from the breakdown of
isostaticity at the jamming transition point, as predicted
in~\cite{WNW05,Wy12}. Furthermore, the regularity of these distributions
leads to a trivial linear distance dependence of the physical quantities
near the jamming transition point.  We find that the critical exponents
change discontinuously at zero asphericity, meaning that an
infinitesimal asphericity is enough to alter the universality of the
jamming transition.

The paper is organized as follows.  In Sec.~\ref{154708_12Feb19}, we
introduce the model and in Sec.~\ref{154729_12Feb19}, we derive the free
energy and saddle point equations with the replica method.  In
Sec.~\ref{154821_12Feb19}, we calculate the phase diagram using the
replica symmetric ansatz.  In Sec.~\ref{155003_12Feb19}, we argue the
scaling behavior of the model above and at the jamming transition point
and in Sec.~\ref{155133_12Feb19}, we calculate the density of states in
the UNSAT(jammed) phase and compare it with previous numerical results
on ellipsoids.  Finally in Sec.~\ref{155216_12Feb19}, we summarize our
results and give some perspectives. We discuss the technical details in
the appendix.


\section{The polydisperse perceptron model}
\label{154708_12Feb19} 

Jamming of particles can be regarded as a special case of a more
general class of problems where one needs to find an assignement
of a large set of continuous variables to satisfy a set of constraints \cite{franz2016simplest}.
These problems are therefore called Continuous Constraint Satisfaction Problems (CCSP).
In this general setting, there is a precise dictionary between jamming
of spheres and CCSP that allows to identify several physical quantities, from the pressure
to gap variables and number of contacts, in both class of problems. 
Here we will not review this dictionary extensively but we will limit ourselves 
to discuss it online while presenting the application of this line of reasoning
to study jamming of non-spherical particles. The interested reader can find more
details in \cite{franz2015, franz2017}.

The perceptron is a well known and studied linear classifier in machine
learning~\cite{rosenblatt1958perceptron,Ga88,gardner1988optimal,nishimori2001}. Here,
following \cite{franz2016simplest}, we turn it into a continuous
constraint satisfaction/optimization problem to study the jamming
transition. The model studied in \cite{franz2016simplest} is defined in
terms of a state vector $\bx=\{x_1,\ldots, x_N\}$ which is
$N$-dimensional and lives on the $(N-1)$-dimensional hypersphere defined
by $|\bx|^2=N$.  Furthermore one defines a set of $M=\alpha N$
\emph{quenched} $N$-dimensional random vectors $\{\bxi^\mu =
\{\xi^\mu_1,\ldots, \xi^\mu_N\}\}_{\mu=1,\ldots, M}$ whose components
are Gaussian random variables with zero mean and unit variance.  Given
each random vector (often referred to as ``pattern'') and the state
vector of the system, one can define a gap variable as
\begin{equation}
h_\mu= \frac{\bx\cdot\bxi^\mu}{\sqrt{N}}-\sigma.\label{120607_14Feb19}
\end{equation}
where $\sigma$ is a control parameter.  The constraint satisfaction
problem is defined by asking for a vector $\bx$ that satisfies all the
constraints
\begin{equation}
h_\mu\geq 0 \ \ \ \ \ \forall \mu=1,\ldots, M\:.
\end{equation}
On general grounds one can expect that if $\alpha$ is small enough it is
possible to find a satisfiable (SAT) configuration of $\bx$ while if
$\alpha$ is large enough this is not possible and the problem becomes
unsatisfiable (UNSAT). Indeed, in the thermodynamic limit $N\to \infty$,
there exists a sharp SAT/UNSAT phase transition that marks the boundary
between the two phases.  The parameter $\sigma$ controls the convexity
of the problem: if $\sigma>0$ the constraint satisfaction problem is
convex while if $\sigma<0$ it is non-convex.  In
\cite{franz2016simplest} it has been shown that for $\sigma<0$ the
problem has a replica symmetry breaking phase near the jamming point as
hard spheres in infinite dimensions \cite{kurchan2013exact}, and the
SAT/UNSAT transition is in the same universality class of jamming of
spherical particles.

In this work, the perceptron model is extended to describe the jamming
of nonspherical particles. We consider the polydisperse perceptron model
that has been introduced in~\cite{brito2018universality} and here we
first recall its definition and then we construct its full analytical
solution. For non-spherical particles like ellipsoids, the relative
distance between constituent particles depends on their relative
angles. Therefore the actual gap between particles is a function of both
the distance of their centers and relative orientation.  
More precisely, consider non-spherical particles, with $\Delta$ being the linear deviation from spherical shape
(for example, in ellipsoids $\Delta$ is the deviation of the aspect ratio from 1) and
with the asphericity 
parameter being
\begin{equation}
\mathcal{A}(\Delta) = \frac{S_d(\Delta)}{V_d(\Delta)^{\frac{d-1}{d}}}
 \frac{V_d(0)^{\frac{d-1}{d}}}{S_d(0)}-1 \ ,
 \end{equation}
where $S_d(\Delta)$ and $V_d(\Delta)$ represent the surface and volume
of a $d$-dimensional particle, respectively.  Note that because the function
$\mathcal{A}(\Delta)$ has a minimum at $\Delta=0$, it can be expanded as
$\mathcal{A}(\Delta) = \mathcal{A}''(0)\Delta^2/2 + \cdots$, which leads
to $\Delta \sim \mathcal{A}^{1/2}$.
Note also that
for small $\Delta$, the gap between two particles can be written
as~\cite{brito2018universality} 
\begin{equation}\label{eq:gapscal1}
h(r,\theta) \sim h(r) + \Delta \, f(\theta) \ ,
\end{equation}
 where $r$ is the distance between the centers,
$h(r) = r - \sigma$ is the gap for spherical particles of diameter $\sigma$, and $f(\theta)$ is some function
of the non-spherical degrees of freedom.

One can think about the
orientation of a particle as a dynamical \emph{internal} degree of
freedom that each particle can change in order to satisfy better the
constraints.  In this way one can construct more general models in which
beyond the standard degrees of freedom that are the position of the
particles, one can consider internal degrees of freedom that must have
the same role as orientations for non-spherical particles.  
In practice, any internal degree of freedom such that Eq.~\eqref{eq:gapscal1} holds should lead to similar results.
 For example, in~\cite{brito2018universality} we have shown that in the small
asphericity limit, the interaction potential of ellipsoids maps to the
interaction potential of a model of \emph{breathing} spheres, namely
spheres that can change their diameter~\cite{BLW18}.
This suggests that one can extend the perceptron model studied in
Ref.~\cite{franz2016simplest,franz2017}, by replacing the effective
diameter $\sigma$ by a fluctuating one \cite{brito2018universality}
\begin{equation}
\sigma\to \sigma + \Delta R_\mu \ ,
\end{equation}
so that the new gap variables become
\begin{equation}
 h_\mu = \frac{\bx\cdot\bxi^\mu}{\sqrt{N}}-\sigma - \Delta R_\mu\:.\label{122315_14Feb19}
\end{equation}
The variables $\{R_\mu\}_{\mu=1,\ldots M}$ are additional internal
dynamical variables that can be used to find solutions to the constraint
satisfaction problem. 
We bound their amplitude by enforcing them to
live on a $(M-1)$-dimensional hypersphere
\begin{equation}
|\bm{R}|^2\equiv \sum_{\mu=1}^M R_\mu^2=M\:.
\label{sphericalR}
\end{equation}
Because Eq.~\eqref{122315_14Feb19} has the same structure of
Eq.~\eqref{eq:gapscal1}, we expect this modified perceptron to fall into
the same universality class of non-spherical particles with $\mathcal{A}
\sim \Delta^{1/2}$.  The control parameter $\Delta$ thus tunes the asphericity
of the problem: in the limit $\Delta\to 0$, the model reduces to the
standard perceptron model, which corresponds to the system consisting of
spherical particles, except at the jamming transition point. However at
the jamming transition point, the model leads to the completely
different scaling from the standard perceptron even in the $\Delta\to 0$
limit, as we discuss in the rest of the manuscript.


In order to solve the constraint satisfaction problem defined by the
polydisperse perceptron model we can define a cost function
\begin{align}
 U_N &= \sum_{\mu=1}^M v(h_\mu) \ .
\end{align}
The local potential $v(h_\mu)$ has to satisfy the property that
\begin{equation}
v(h) \begin{cases}
=0 & h\geq 0\\
>0 & h<0\:.
\end{cases}
\end{equation}
The choice that we adopt here is the one describing harmonic particles
$v(h) = h^2\theta(-h)/2$.  The particular choice of the cost function
does not affect the SAT phase as well as the SAT/UNSAT threshold, while
it deeply affects the properties of the jammed phase \cite{franz2017,
franz2019critical}.  In order to impose the spherical constraint of
Eq.~(\ref{sphericalR}) we add a chemical potential $\mu_N$ to the cost
$U_N$ so that the final interaction potential reads
\begin{align}
 V_N = U_N + \mu_N,\qquad U_N = \sum_{\mu=1}^M v(h_\mu),\qquad \mu_N
 = \frac{k_R}{2}\sum_{\mu=1}^M R_\mu^2
\end{align}
where $k_R$ is a Lagrange parameter used to enforce
Eq.~(\ref{sphericalR}).  In the next section we study the partition
function of the model at inverse temperature $\beta=1/T$ defined as
\begin{equation}
Z=\int_{|\bx|^2=N} \de \bx \int_{|\bm{R}|^2=M} \de \bm{R} \, e^{-\beta V_N}
\end{equation}
and we study the corresponding average free energy using the replica method \cite{mezard1987}.

\section{Free energy and thermodynamic quantities}
\label{154729_12Feb19} The free energy of the model can be calculated by
using the replica trick:
\begin{align}
 -\beta f = \lim_{n\to 0}\frac{\log \overline{Z^n}}{nN},\label{113422_14Jan19}
\end{align}
where the overline denotes the average over the disorder
$\{\bxi^\mu\}_{\mu=1,\ldots, M}$. As usual for integer $n$, one
interprets the $n$th power of the partition function as the partition
function of a system of $n$ copies or replicas of the original system,
all subjected to the same realization of disorder. In this way, using
standard manipulations, the free energy of the replicated system can be
obtained as {\medmuskip=0mu \thinmuskip=0mu \thickmuskip=0mu
\begin{align}
-\beta n f  \equiv \frac{\log \overline{Z^n}}{N} \approx \frac{1}{2}\log \det Q
 -n\frac{\alpha}{2}\log \tk + \alpha \log \left(e^{\frac{1}{2}\sum_{ab}Q_{ab}\pdiff{^2}{h_a\partial h_b}}
\left.
 \prod_{a=1}^n
 e^{-\beta\phi_{\rm eff}(h_a)}\right|_{h_a=0}
 \right),\label{113413_14Jan19}
\end{align}}
where $\tk = \beta k_R$, and $Q$ denotes the $n\times n$ overlap matrix
$Q_{ab}= \ave{\bx^a\cdot\bx^b}/N$, $a,b = 1,\cdots, n$ that has diagonal
elements equal to one due to the spherical constraint on $\bx$ while its
off diagonal elements are obtained as a saddle point of
Eq.~(\ref{113413_14Jan19}). In Eq.~(\ref{113413_14Jan19}) we have also
introduced the effective potential
\begin{align}
 e^{-\beta \phi_{\rm eff}(r)} = \gamma_{\Delta^2/\tk}*e^{-\beta v(r)}.
\end{align}
Here $\gamma_A(x)$ denotes a Gaussian function of zero mean and variance
$A$, and the star product denotes the convolution, $f*g(x) =
\int_{-\infty}^\infty \de y f(y)g(x-y)$.  The parameter $k_R$ should
also be determined by a saddle point of Eq.~(\ref{113413_14Jan19}), in
order to enforce the constraint in Eq.~\eqref{sphericalR}.  The saddle
point equations for the overlap matrix $Q$ can be solved only assuming
some ansatz that allows one to take the analytic continuation $n\to
0$. Following the standard strategy of replica theory, we assume a
hierarchical ansatz for $Q$~\cite{mezard1987spin}, in which $Q$ is
encoded by a continuous function $q(x)$ defined for $x\in [0,1]$ with
the boundary conditions $q(x)=q_m$ for $x\in [0,x_m)$ and $q(x)=q_M$ for
$x\in (x_M,1]$.  It is practical to define $x(q)$ as the inverse
function of $q(x)$ for $x\in [x_m,x_M]$. After a standard but slightly
lengthy procedure (see \ref{124714_14Feb19} and \cite{franz2017}), one
obtains the saddle point condition for $x(q)$ as the solution of:
\begin{align}
 \frac{q_m}{\lambda(q_m)^2} +
 \int_{q_m}^q \de p \frac{1}{\lambda(p)^2}
 &= \alpha \int \de h P(q,h)f'(q,h)^2 \ ,\label{104920_13Mar18}
\end{align}
where
\begin{align}
 \lambda(q) &= 1-q_M + \int_q^{q_M}\de p x(p) \ .
\end{align}
 In Eq.~(\ref{104920_13Mar18}), $f(q,h)$ and $P(q,h)$ are the solutions of the Parisi
equations
\begin{align}
 \dot{f}(q,h) &= -\frac{1}{2}\left[f''(q,h)+x(q)f'(q,h)^2\right],\new
 \dot{P}(q,h) &= \frac{1}{2}\left[P''(q,h)-2x(q)\left(P(q,h)f'(q,h)\right)'\right],\label{190748_18Jan19}
\end{align}
where $\dot{g}(q,h) = \partial g(q,h)/\partial q$ and $g'(q,h)=\partial
g(q,h)/\partial h$, and the boundary conditions are
\begin{align}
 f(q_M,h) &= \log\gamma_{1-q_M+\Delta^2/\tk}*e^{-\beta v(h)} \ ,\new
 P(q_m,h) &= \gamma_{q_m}(h+\sigma) \ .\label{153156_18Jan19}
\end{align}
In the continuous RSB phase, $x(q)$ becomes a monotonic function of $q$,
suggesting that Eq.~(\ref{104920_13Mar18}) can be differentiated
w.r.t. $q$ \cite{franz2017, sommers1984distribution}.  The first and
second derivatives lead to
\begin{align}
 \frac{1}{\lambda(q)^2} &= \alpha \int \de h
 P(q,h)f''(q,h)^2,\label{184453_18Jan19}\\
 x(q)  &= \frac{\lambda(q)}{2}\frac{\int \de h P(q,h)f'''(q,h)^2}{\int \de h P(q,h)\left[ f''(q,h)^2 + \lambda(q)f''(q,h)^3\right]}.\label{191922_18Jan19}
\end{align}
Finally, the spherical constraint for $R_\mu$,
Eq.~(\ref{sphericalR}), reduces to
\begin{align}
 1 &= \frac{1}{\tk} + \frac{\Delta^2}{\tk^2}\int \de h P(q_M,h)\left[f''(q_M,h)
 + (f'(q_M,h))^2\right].\label{105449_13Mar18}
\end{align}
Eqs.~(\ref{104920_13Mar18}-\ref{105449_13Mar18}) represent the set of
saddle point equations through which we can obtain the free energy of
the model.

For later convenience, we introduce several thermodynamic quantities
that we shall discuss in this manuscript. First we define the gap
distribution as the following Boltzmann average
\begin{align}
\rho(h)  = \frac{1}{M}\ave{\sum_{\mu=1}^M \delta(h-h_\mu)},
\end{align}
which can be calculated from the first derivative of the free energy
w.r.t. the interaction potential~\cite{franz2017}:
\begin{align}
 \rho(h) = \frac{1}{\alpha}\fdiff{f}{v(h)} = e^{-\beta v(h)}
 \int \de z P(q_M,z)e^{-f(q_M,z)}\gamma_{1-q_M+\Delta^2/k_R}(z-h) \ .\label{175504_19Jan19}
\end{align}
Using $\rho(h)$, one can calculate the isostaticity index as
\begin{align}
 z &\equiv \frac{1}{N}\ave{\sum_{\mu=1}^M \theta(-h_\mu)} = \alpha \int \de h\rho(h)\theta(-h).\label{011940_22Jan19}
\end{align}
The value of $z$ equals one when the system is isostatic, namely,
when the number of UNSAT gaps is equal to the dimension of $\bx$. Note that the normalization of Eq.~(\ref{011940_22Jan19})
is $N$, which is not the total number of degrees of freedom given by
$N+M$ \footnote{
The reason for this normalization is the following. One can easily show~\cite{WNW05,franz2015,brito2018universality} that the matrix of second derivatives
of $U_N$ has precisely $N \max(1-z,0)$ zero modes. For $\Delta=0$, i.e. in absence of $\mu_N$, stability then requires $z\geq 1$. 
Jamming has the minimal number of constraints and is then isostatic,
$z=1$, corresponding to a number of constraints precisely equal to the number of degrees of freedom~\cite{WNW05}.
As soon as $\Delta>0$, one has $M+N$ degrees of freedom, and the matrix of second derivatives of $U_N$ has then $N \max(\alpha+1-z,0)$ zero modes.
Isostaticity would then correspond to $z=1+\alpha$, but
the term $\mu_N$ can stabilize some of the zero modes of $U_N$. 
Hence, the system is not constrained to be isostatic at jamming.
We will show analytically that when $\Delta$ is small $0<z-1\sim \Delta^{1/2}<\alpha$ at the non-convex jamming point, 
and therefore the system is hypostatic (even if in our notation $z>1$).}. 
The pressure $p$ is also calculated from $\rho(h)$:
\begin{align}
 p &\equiv -\frac{1}{N}\ave{\sum_{\mu=1}^M h_\mu\theta(-h_\mu)} = -\alpha \int \de h \rho(h)h\theta(-h).\label{183458_15Feb19}
\end{align}
For comparison with numerical results, we introduce the positive gap distribution
\begin{align}
 g(h) &\equiv \theta(h)\frac{\rho(h)}{\int_0^\infty \de h \rho(h)},\label{003451_22Jan19}
\end{align}
and the force distribution
\begin{align}
 P(f) &\equiv \theta(-h)\frac{\rho(h)\diff{h}{f}}{\int_{-\infty}^0 \rho(h)\diff{h}{f}\de f},\label{003504_22Jan19}
\end{align}
where $f= -h/p$ for negative $h$.

\section{Phase diagram}
\label{154821_12Feb19}

 \subsection{Replica symmetric jamming transition point}
\begin{figure}
\centering \includegraphics[width=.8\textwidth]{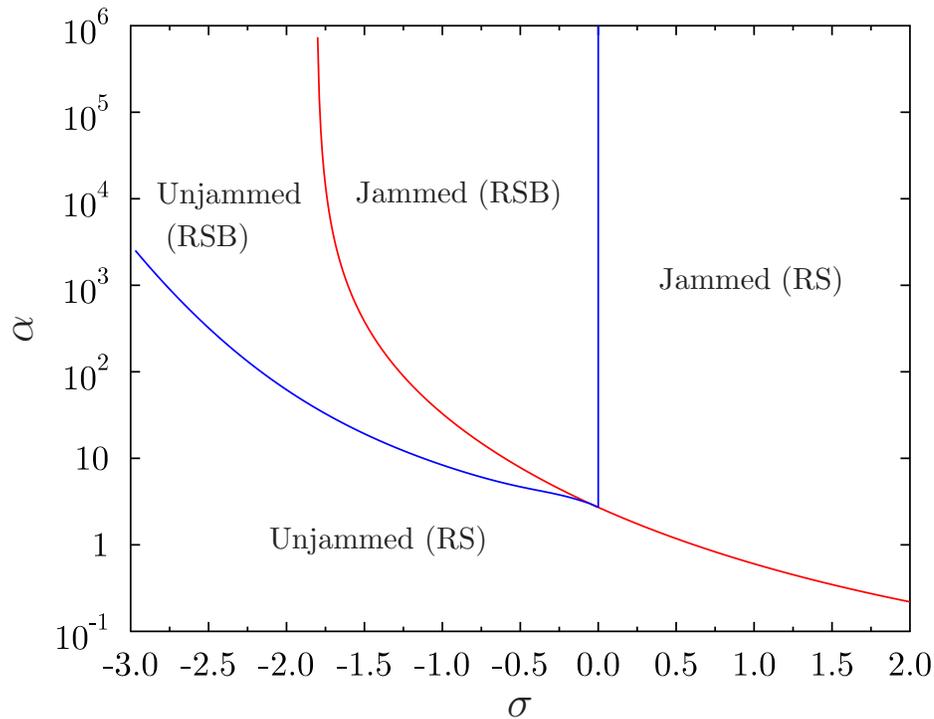}
\caption{Phase diagram of the perceptron for $\Delta=0.1$. The red line
denotes the jamming transition point computed using a replica symmetric
ansatz.  The blue lines denote the RS-RSB transition. Picture taken from
Ref.~\cite{brito2018universality}.}  \label{163148_15Jan19}
\end{figure}

We first consider the simplest ansatz for $Q$, namely the replica symmetric (RS) form  $Q_{ab}=\delta_{ab} +
(1-\delta_{ab})q$. In this case the corresponding saddle point equations reduce to
\begin{align}
 \frac{q}{(1-q)^2} = \alpha \int \de h P_{\rm RS}(q,h)f_{\rm RS}'(q,h)^2 \ ,\label{020701_17Jan19}
\end{align}
where
\begin{align}
 f_{\rm RS}(q,h) &= \log \gamma_{1-q+\Delta^2/\tk}*e^{-\beta v(h)},\new
 P_{\rm RS}(q,h) &= \gamma_{q}(h+\sigma)\:.\label{161951_15Feb19}
\end{align}
We want to consider what happens in the zero temperature limit where the
partition function gives either the Gardner volume of solutions of the
constraint satisfaction problem \cite{Ga88, GD88} in the SAT phase or
the ground state energy in the UNSAT phase.  In this second case the
overlap $q$ parametrizes the typical overlap between two configurations
in the ground state basin and therefore for $T\to 0$ we have that $q\to
1$. Therefore we can expand $q$ as \cite{franz2017}
\begin{align}
 q \approx 1- T\chi,\label{170211_14Jan19}
\end{align}
where $\chi$ is a constant. At the jamming transition point, $\chi$
diverges to infinity since in the SAT phase $q<1$ \cite{franz2017}.  For
$\beta\gg 1$, we obtain (see Appendix~C in Ref.~\cite{franz2017})
\begin{align}
 f_{\rm RS}(q,h) \sim -\beta \frac{h^2}{2(1+\chi+\Delta^2/k_R)}\theta(-h).
\end{align}
Substituting this into Eq.~(\ref{020701_17Jan19}), we obtain
\begin{align}
 \left(1+\frac{1}{\chi}+ \frac{\Delta^2}{k_R\chi}\right)^2
 = \alpha G(\sigma)\label{192602_14Jan19},
\end{align}
where we have introduced an auxiliary function:
\begin{align}
 G(\sigma) = \int_{-\infty}^0 dh \gamma_1(h+\sigma)h^2.
\end{align}
With a similar calculation, one can show that 
Eq.~(\ref{105449_13Mar18}) reduces to
\begin{align}
1 = \frac{\Delta^2}{(\Delta^2+k_R(1+\chi))^2}G(\sigma).\label{192606_14Jan19}
\end{align}
Eqs.~(\ref{192602_14Jan19}) and (\ref{192606_14Jan19}) can be solved for $k_R$,
\begin{align}
 k_R = \frac{\Delta}{\chi \sqrt{\alpha}},
\end{align}
which implies that $k_R$ vanishes on approaching the jamming
transition point as $k_R\sim \chi^{-1}$.  Substituting this into
Eq.~(\ref{192602_14Jan19}) and taking the $\chi\to\infty$ limit, we obtain
the jamming transition point $\alpha_J$:
\begin{align}
\alpha_J(\sigma,\Delta) &= \left(\frac{1}{\sqrt{G(\sigma)}-\Delta}\right)^2.\label{161700_15Jan19}
\end{align}
The same equation is obtained by investigating the saddle point
equations in the unjammed phase, see \ref{173919_15Feb19}. In
Fig.~\ref{163148_15Jan19}, we show the typical behavior of $\alpha_J$
calculated by Eq.~(\ref{161700_15Jan19}) for $\Delta=0.1$.
$\alpha_J$ increases with decreasing $\sigma$, which is a reasonable
result considering that $\sigma$ corresponds to the effective diameter.
For fixed $\sigma$, $\alpha_J$ increases as $\Delta$ increases, which is
consistent with the numerical results of nonspherical particles of small
asphericity~\cite{williams2003,donev2004,jiao2010,werf2018}.  On the
contrary, for larger asphericity, the jamming density of a particle
system exhibits a non-monotonic
behavior~\cite{williams2003,donev2004,jiao2010,werf2018}, which is
not captured by this extended perceptron model, which only represents
nonspherical particles with small asphericity up to
$O(\Delta^2)=O(\mathcal{A})$~\cite{brito2018universality}.

\subsection{RS-RSB phase boundary}
The RS ansatz  breaks
down in the (full) replica symmetric breaking (RSB) phase where $q(x)$
becomes a continuous function of $x$. In this case, the inverse function
$x(q)$ is also a continuous function of $q$.  Thus, at the RS-RSB phase
boundary, $q$ calculated with the RS ansatz should satisfy
Eq.~(\ref{184453_18Jan19}), which leads to
\begin{align}
\left(1+\frac{1}{\chi}+\frac{\Delta^2}{k_R\chi}\right)^2
 = \alpha\int_{-\infty}^0 dh \gamma_1(h+\sigma).\label{171250_15Jan19}
\end{align}
Using Eqs.~(\ref{192602_14Jan19}) and (\ref{171250_15Jan19}), it can be
shown that the RS-RSB boundary in the jammed phase is the vertical line
defined by $\sigma=0$, see the blue vertical line in
Fig.~\ref{163148_15Jan19}. Similarly, one can calculate the RS-RSB phase
boundary by substituting the RS result in the unjammed phase into
Eq.~(\ref{184453_18Jan19}), see the blue line in the unjammed phase of
Fig.~\ref{163148_15Jan19}.  The resulting figure shows that the jamming
transition line lies in the RSB region when $\sigma<0$, as in the case
of the standard spherical perceptron~\cite{franz2016simplest}.

\section{Jamming scaling in the RSB phase}
\label{155003_12Feb19}

In this section, we discuss the scaling behavior of the jamming
transition for $\sigma<0$, where jamming is described by the RSB
equations and belongs to the same universality class of particle
systems. For the sake of brevity, here we only discuss the scaling in
the jammed phase, but the scaling solutions in the unjammed phase can be
constructed in a very similar way, see \ref{181017_15Feb19} for the
details.

\subsection{Scaling solutions in the zero temperature limit}
First we derive the asymptotic forms of the relevant equations in the
zero temperature limit, $T\to 0$. As in the case of the RS analysis, we
expand as $q_M = 1-T\chi$, where $\chi$ has a finite limit for $T\to 0$ and
diverges at the jamming transition point. We introduce the following
scaling variables for $\chi\gg 1$~\cite{franz2017}:
\begin{align}
 K &= \frac{\chi k_R}{\Delta^2},& y(q) &= \frac{\beta}{\chi}x(q),& \pi(q) &= 1 + \frac{1}{K}
 + \int_q^1 dp y(p),& m(q,h)&= \pi(q)f'(q,h) \ .\label{184437_18Jan19}
\end{align}
Then, the gap distribution $\rho(h)$, Eq.~(\ref{175504_19Jan19}), is 
\begin{align}
 \rho(h) \sim \theta(h)P(1,h) + \theta(-h)(1+\chi+\Delta^2/k_R) P(1,(1+\chi+\Delta^2/k_R)h).\label{181802_18Jan19}
\end{align}
Using Eq.~(\ref{011940_22Jan19}) and (\ref{184453_18Jan19}), 
we obtain 
\begin{align}
 z &= \left(1 + \frac{1}{\chi} + \frac{1}{K}\right)^2.\label{145322_23Jan19}
\end{align}
Away from the jamming point, it has been shown that $y(q)$
is described by the following scaling solution near $q\sim
1$~\cite{franz2017}:
\begin{align}
 y(q) &\sim \frac{\Y}{\sqrt{1-q}},\label{172505_19Jan19}
 \end{align}
which we refer to as the ``regular'' solution.  For $m(q,h)$, we assume
\begin{align}
m(q,h) &\sim -\frac{\chi(1+K^{-1})}{1+\chi(1+ K^{-1})}\sqrt{1-q}\M\left(\frac{h}{\sqrt{1-q}}\right).
 \label{193732_18Jan19}
\end{align}
Using Eq.~(\ref{191922_18Jan19}), one can calculate $\Y$ as
\begin{align}
 \Y &\sim \frac{K+\chi K + \chi}{2(K+\chi)}
 \frac{P(1,0)\int dt \M''(t)^2}{\int_{-\infty}^0 dh P(1,h)}.\label{173049_19Jan19}
\end{align}

\subsection{Jamming for $\Delta=0$}

For self-completeness, we first review the scaling behavior of the
standard perceptron model ($\Delta=0$), which corresponds to the jamming
of spherical particles and has been already well investigated in previous
work~\cite{franz2017}.  In this case, the isostaticity index defined in
Eq.~(\ref{145322_23Jan19}) reduces to~\cite{franz2017}
\begin{align}
 z &= \left(1 + \frac{1}{\chi}\right)^2.\label{001908_22Jan19}
\end{align}
At the jamming transition point, $\chi\to\infty$, one obtains
\begin{align}
 z \to 1,\label{001913_22Jan19}
\end{align}
meaning that the system becomes isostatic: the number of contacts is the same
of that of the degree of freedom.  Away from the jamming
transition point, $z$ increases as
\begin{align}
\delta z &\equiv z-1 = \frac{1}{\chi}.
\end{align}
For $\Delta=0$, Eq.~(\ref{173049_19Jan19}) reduces to
\begin{align}
 \Y &\sim \frac{1+\chi}{2}
 \frac{P(1,0)\int dt \M''(t)^2}{\int_{-\infty}^0 dh P(1,h)}\label{192612_20Jan19}
\end{align}
and for $\chi\gg 1$, we get 
\begin{align}
\Y &\sim \chi P(1,0) \sim \delta z^{-1}P(1,0),
\end{align}
implying that the scaling solution Eq.~(\ref{172505_19Jan19}) breaks
down when the system becomes isostatic and $\delta z=0$. Indeed in the
isostatic limit, the scaling ansatz for $y(q)$ becomes:
\begin{align}
 y(q) \sim \frac{y_J}{(1-q)^{1/\kappa}},\label{182717_15Feb19}
\end{align}
where the value of the critical exponent $\kappa=1.41574\ldots$ is
obtained by solving the Parisi equations in the scaling
regime~\cite{franz2017}.  Hereafter, we shall refer to
Eq.~(\ref{182717_15Feb19}) as the ``critical'' scaling solution. The
matching argument between the regular and critical scaling solutions
determines the asymptotic behavior of $P(1,h)$ as~\cite{franz2017}
\begin{align}
 P(1,h) \sim \begin{cases}
	      \chi p_{-}(h\chi) & {\rm for}\ h\sim -\chi^{-1},\\
	      \chi^{\gamma\psi}p_0(h\chi^{\psi})
	      & {\rm for}\ \abs{h}\sim \chi^{-\psi},\\
	      p_+(h) & {\rm for}\ h\gg \chi^{-\psi},
	     \end{cases}\label{185637_20Jan19}
\end{align}
where 
\begin{align}
 p_+(t\to 0^+)&\sim t^{-\gamma},&
 p_0(t\to\infty)&\sim t^{-\gamma},&
 p_{-}(t\to 0^{-})&\sim \abs{t}^\theta,&
  p_0(t\to-\infty)&\sim \abs{t}^\theta.
\end{align}
The critical exponents are
\begin{align}
 \gamma &= \frac{2-\kappa}{\kappa}=0.413,&
 \theta &= \frac{3\kappa-4}{2-\kappa}= 0.421,&
 \psi &= \frac{\kappa}{2(\kappa-1)} = 1.703.
\end{align}
Using Eqs.~(\ref{183458_15Feb19}), (\ref{181802_18Jan19}) and
(\ref{185637_20Jan19}), the pressure is calculated
as 
\begin{align}
 p &= -\alpha\int_{-\infty}^0 dh P(1,h) h \sim -\delta z^2 \int_{-\infty}^0 dt p_{-}(t)t,
\end{align}
leading to
\begin{align}
 \delta z \sim p^{1/2}.\label{003640_22Jan19}
\end{align}
This is consistent with the numerical results of the jamming transition
of spherical particles interacting with the harmonic
potential~\cite{ohern2003}. Using Eqs.~(\ref{003451_22Jan19}),
(\ref{003504_22Jan19}), (\ref{181802_18Jan19}), and
(\ref{185637_20Jan19}), we arrive at the asymptotic forms of the positive
gap and force distributions:
\begin{align}
 g(h) &\sim \begin{cases}
	     \delta z^{-\psi\gamma} p_0(h\delta z^{-\psi}) & (h\sim \delta z^{\psi}),\\
	     h^{-\gamma} & (h\sim 1)
	    \end{cases},\new
P(f) &\sim \begin{cases}
	    \delta z^{\theta\omega}p_0(f\delta z^{-\omega}) & (f\sim \delta z^{\omega})\\
	    f^\theta & (f\sim 1)
	   \end{cases},
 \label{190801_20Jan19}
\end{align}
where 
\begin{align}
 \omega &= \psi-1 = 0.703.
\end{align}
Eqs.~(\ref{190801_20Jan19}) show that the gap and force distributions
exhibit a power law behavior if the system is isostatic, $\delta z=0$, while
they remain finite and regular if the system is not isostatic, $\delta
z\neq 0$.

\subsection{Jamming for $\Delta>0$}
We now discuss the scaling behavior for $\Delta>0$, corresponding to the
jamming of nonspherical particles. We first study the equations in
the jamming limit, $\chi\to\infty$. The isostaticity index at the
jamming transition point is
\begin{align}
 z = \left(1 + \frac{1}{K}\right)^2,\label{192540_20Jan19}
\end{align}
implying that
\begin{align}
 \delta z \sim \frac{1}{K}\neq 0.\label{133952_23Jan19}
\end{align}
Eq.~(\ref{173049_19Jan19}) at the jamming point becomes
\begin{align}
 \Y &\sim \frac{1+K}{2}\frac{P(1,0)\int dt \M''(t)^2}{\int_{-\infty}^0 dhP(1,h)}.\label{192622_20Jan19}
\end{align}
Note that, unlike for jamming of spherical particles, $\Y\sim K
P(1,0)\sim \delta z^{-1} P(1,0)$ remains finite even at jamming, implying that the regular scaling solution,
described by Eq.~(\ref{172505_19Jan19}) persists even at the jamming transition. The regular solution connects to the critical solution described by
Eq.~(\ref{182717_15Feb19}), in the spherical limit $\Delta\to 0$.  It is
worth noting that Eqs.~(\ref{001908_22Jan19}) and (\ref{192612_20Jan19})
can be identified with Eqs.~(\ref{192540_20Jan19}) and
(\ref{192622_20Jan19}), if one replaces $\chi$ with $K$, implying that
the scaling form of $P(1,h)$ for $K\gg 1$ is obtained by simply
replacing $\chi$ in Eq.~(\ref{185637_20Jan19}) with $K$ :
\begin{align}
 P(1,h) \sim \begin{cases}
	      K p_{-}(hK) & {\rm for}\ h\sim -K^{-1}\ ,\\
	      K^{\gamma \psi}p_0(h K^{\psi})
	      & {\rm for}\ \abs{h}\sim K^{-\psi}\ ,\\
	      p_+(h) & {\rm for}\ h\gg K^{-\psi} \ .
	     \end{cases}\label{153256_23Jan19}
\end{align}
Using this scaling form Eq.~(\ref{105449_13Mar18}) reduces to
\begin{align}
 \Delta^2 \sim \frac{1}{(1+K)^2}\int_{-\infty}^0 dh P(1,h)h^2
\sim
 \frac{1}{K^2(1+K)^2} \int_{-\infty}^0 dt p_{-}(t)t^2,
\end{align}
which, for $\Delta\ll 1$, leads to
\begin{align}
 K \sim \Delta^{-1/2}.\label{144728_21Jan19}
\end{align}
From this equation and Eq.~(\ref{133952_23Jan19}), we have 
\begin{align}
 \delta z \sim \Delta^{1/2}\to z_J = 1+ c\Delta^{1/2},
\end{align}
which agrees with numerical results for jamming of nonspherical
particles~\cite{mailman2009,werf2018} if one identifies $\Delta$ with
the square root of the asphericity~\cite{brito2018universality}.

\begin{figure}
\centering \includegraphics[width=.4\textwidth]{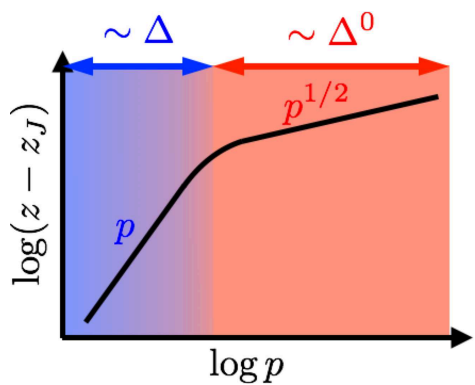} \caption{
 Schematic behavior of the contact number.  Near the jamming transition
 point, $z-z_J\propto p$, while $z-z_J\propto p^{1/2}$ away from
 jamming.}  \label{154005_30Mar19}
\end{figure}

Next, we discuss the scaling behavior above the jamming transition
point, where $\chi$ is large but finite. The pressure can be represented
using $\chi$ and $K$ as
\begin{align}
 p & = -\alpha \int_{-\infty}^0 dh \rho(h)h \sim -\frac{1}{\left(1 + \chi(1+K^{-1})\right)K}\int_{-\infty}^0 dt p_{-}(t)t
 \sim \frac{1}{\chi K},
\end{align}
implying 
\begin{align}
 \frac{1}{\chi}\sim Kp\sim \frac{p}{\Delta^{1/2}}.\label{151854_23Jan19}
\end{align}
Then, the isostaticity index, Eq.~(\ref{145322_23Jan19}),
can be expanded as
\begin{align}
 \delta z = z-1 &\sim \frac{1}{K} + \frac{1}{\chi} \sim c_1\Delta^{1/2} + c_2 \frac{p}{\Delta^{1/2}},\label{163152_16Feb19}
\end{align}
where $c_1$ and $c_2$ are constants. Note that $z$ linearly depends on
$p$. This is consistent with numerical results for the jamming
transition in ellipsoids~\cite{schreck2010} and in marked contrast with
the standard perceptron, corresponding to spherical particles, where
$\delta z \sim p^{1/2}$ according to Eq.~(\ref{003640_22Jan19}). Note
that $\delta z$ should converge to Eq.~(\ref{003640_22Jan19}) in the
$\Delta\to 0$ limit, which requires the following scaling form:
\begin{align}
 \delta z &= \Delta^{1/2}\Z(\Delta^{-1}p)\label{zscaling},
\end{align}
where $\Z(x)\to \rm{const}$ for $x\ll 1$ and $\Z(x)\to x^{1/2}$ for
$x\gg 1$~\cite{brito2018universality}. Eq.~(\ref{zscaling}) implies
that, upon approaching the jamming transition point, the scaling
behavior of $z-z_J$ is changed qualitatively at $p\sim \Delta$ from the
scaling of spherical particles $z-z_J\propto p^{1/2}$ to that of
nonspherical particles $z-z_J\propto p$, as shown in
Fig.~\ref{154005_30Mar19}.
From Eqs.~(\ref{184437_18Jan19}),
(\ref{144728_21Jan19}), and (\ref{151854_23Jan19}), we have
\begin{align}
 k_R &= \Delta^2\frac{K}{\chi} \sim \Delta p.\label{164306_1Feb19}
\end{align}
Finally, by substituting Eq.~(\ref{151854_23Jan19}) into
Eq.~(\ref{153256_23Jan19}), we can calculate $g(h)$ and $P(f)$ at the
jamming transition point.  Interestingly, the resulting equations are
identical to Eqs.~(\ref{190801_20Jan19}), meaning that the criticality
of $g(h)$ and $P(f)$ is controlled by $\delta z$ only. Because $\delta
z>0$ for nonspherical particles, $g(h)$ and $P(f)$ remain finite and
regular functions even at the jamming transition point.

\section{Vibrational density of states}
\label{155133_12Feb19}

In this section, we discuss the scaling of the vibrational density of
states $D(\omega)$ of the polydisperse perceptron. We focus on the
region $\sigma<0$ where the model falls into the same universality class
of nonspherical particles, as shown in the previous sections. In the
UNSAT phase the ground state is a configuration of the state vectors
$\bx$ and $\bm{R}$ that minimizes the potential $V_N$.  By defining
$\rho(\lambda)$ as the eigenvalue distribution of the Hessian of $V_N$
computed in the ground state, the density of states is given by
\begin{align}
D(\omega)= 2\omega \rho(\lambda=\omega^2).\label{164048_16Feb19}
\end{align}
To calculate $\rho(\lambda)$, we consider a slightly modified
interaction potential:
\begin{align}
 V_N &= \sum_{\mu=1}^M \frac{h_\mu^2}{2}\theta(-h_\mu)
 + \frac{k_R}{2}\sum_{\mu=1}^M R_\mu^2 
 -\frac{\zeta}{2}\left(\sum_{i=1}^N x_i^2 -N\right), 
 \label{potential_UNSAT}
\end{align}
where $\zeta$ denotes the Lagrange multiplier that is needed to impose
the spherical constraint $|\bx|^2= N$.  In the UNSAT phase, the first
term of Eq.~(\ref{potential_UNSAT}) is a sum over the UNSAT constraints,
which are the $Nz$ contacts defined as the gaps that satisfy
$\theta(-h_\mu)=1$.  For convenience, we hereafter reassign the index
$\mu$ only to these gaps $\mu=1,\cdots Nz$.  Note that in the following, we neglect the
$(\alpha-z)N$ degrees of freedom $R_\mu$ associated to the 
SAT constraints, such that $\theta(-h_\mu)=0$. These degrees of freedom are 
trivially decoupled from the system and, if included, would give rise to a delta function
$\delta(\lambda-k_R)$ in the density of states. Using this convention, we can write a reduced $N(1+z) \times N(1+z)$ Hessian of
the potential $V_N$ as
\begin{align}
H_{ij} &\equiv \pdiff{^2 V_N}{x_i\partial x_j} = \frac{1}{N}\sum_{\mu=1}^{zN}\xi_i^\mu\xi_j^\mu-\zeta\delta_{ij},\new
Q_{\mu\nu}  & \equiv \pdiff{^2 V_N}{R_\mu\partial R_\nu} = \left[\Delta^2+k_R\right]\delta_{\mu\nu},\new
 T_{\mu i} & \equiv \pdiff{^2 V_N}{R_\mu\partial x_i } =-\frac{\Delta \xi_i^\mu}{\sqrt{N}}.
\end{align}
We define the Hessian matrix as
\begin{align}
 \he  &=
 \begin{pmatrix}
  H & T\\
  T^t & Q
 \end{pmatrix}.
\end{align}
One can calculate the eigenvalue distribution $\rho(\lambda)$ of $\he$
by using the Edwards-Jones formula~\cite{edwards1976}:
\begin{align}
 \rho(\lambda) &\equiv \frac{1}{N+zN}\sum_{k=1}^{N+zN}
 \delta(\lambda-\lambda_k)
 = -\frac{2}{(N+zN)\pi}\lim_{\varepsilon\to 0}
 \im\pdiff{}{\lambda}\overline{\log Z(\lambda-i\varepsilon)},\label{174738_29Jan19}
\end{align}
where the overline denotes the average over the quenched disorder
$\xi_i^\mu$, and we have introduced a partition function as 
\begin{align}
 Z(\lambda) &= \int d\bu
 \exp\left[-\frac{1}{2}\bu\cdot(\lambda I_{N+zN}-\he)\cdot\bu\right].
\end{align}
Here $I_d$ denotes the $d$-dimensional unit matrix.  Due to the mean
field nature of the model, we can derive the analytical expression of
$\rho(\lambda)$ (see \ref{160922_16Feb19}):
{\medmuskip=0mu
\thinmuskip=0mu
\thickmuskip=0mu
\begin{align}
 \rho(\lambda) &= \frac{z-1}{1+z}\delta(\lambda-\Delta^2 -k_R)
 + \mathcal{R}(\lambda),\new
 \mathcal{R}(\lambda) &=
 \abs{\frac{\im \sqrt{\prod_{i=1}^4(\lambda-\lambda_i)}}{2\pi(1+z)}
 \left[
 \frac{1}{(\lambda-k_R)(\lambda+\zeta)}
 -\frac{1}{(\lambda-k_R)^2} + \frac{1}{(\lambda-k_R)(\lambda-\Delta^2-k_R)}
 \right]},\label{163127_1Feb19}
\end{align}}
where the explicit expressions of $\lambda_1,\cdots, \lambda_4$ are
given in \ref{160922_16Feb19}. To calculate $\rho(\lambda)$, one needs
to calculate $\zeta$, $z$, and $k_R$ for given $p$ and $\Delta$. In the
RSB phase, there is a useful relation to express $\zeta$ (see \ref{161701_16Feb19}):
\begin{align}
 \zeta &= \frac{1}{1 + \Delta^2/k_R}\left(\sqrt{z}-1\right)^2.
\end{align}
To calculate $z$ and $k_R$, one should solve the RSB equation
numerically, which is a difficult task. Since we are mainly interested in the
scaling property of the density of states, here we assume a suitable
ansatz for $z$ and $k_R$ of the same form discussed in
the previous sections. The scaling of $z$ given by
Eq.~(\ref{zscaling}), can be satisfied by assuming
\begin{align}
\delta z &= \left(c_1 p + c_2 \Delta\right)^{1/2},\label{164355_1Feb19}
\end{align}
where $c_1$ and $c_2$ are constants. To satisfy
Eq.~(\ref{164306_1Feb19}), we assume
\begin{align}
 k_R &= c_3 \Delta p,\label{164400_1Feb19}
\end{align}
where $c_3$ is another constant. Because these constants do not affect the scaling
behavior, we set them to $c_1=c_2=c_3=1$. Substituting
Eqs.~(\ref{164355_1Feb19}) and (\ref{164400_1Feb19}) into
Eq.~(\ref{163127_1Feb19}), and using Eq.~(\ref{164048_16Feb19}), one can
calculate $D(\omega)$.

\begin{figure}
\centering \includegraphics[width=.8\textwidth]{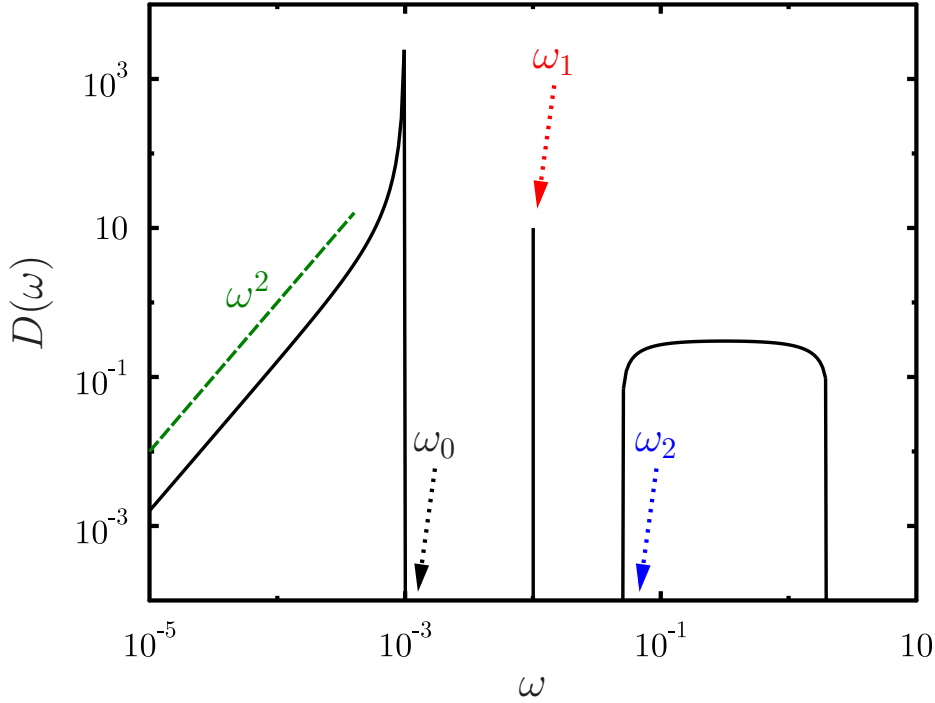} \caption{Density of
 states of the perceptron for $\Delta=10^{-2}$ and $p=10^{-4}$. The black
 solid line denotes the theoretical result.  The green line denotes
 $D(\omega)\sim\omega^2$ scaling. The arrows indicate the characteristic
 frequencies.  } \label{185025_31Jan19}
\end{figure}

In Fig.~\ref{185025_31Jan19}, we show the
typical behavior of $D(\omega)$, which consists of three regions separated by
finite gaps:
\begin{enumerate}
 \item The lowest band is quasi-gapless and ends at $\omega_0$. The
       existence of gapless excitations is a direct consequence of
       RSB as already discussed for the standard
       perceptron~\cite{franz2015}. The weight of the lowest band is
       $f_0\equiv \int_0^{\omega_0^2}d\lambda \rho(\lambda) =
       \int_0^{\omega_0}d\omega D(\omega)= 1/(1+z)$.

\item The delta function at $\omega_1$. The weight is $f_1=(z-1)/(1+z)$.

\item The highest band starting at $\omega_2$. The weight of this band
is $f_2=1/(1+z)$.
\end{enumerate}
Note that $D(\omega)$ is normalized so that $\int_0^\infty d\omega
D(\omega) = f_0 + f_1 + f_2 = 1$. The behavior of $D(\omega)$ that we
find in the polydisperse spherical perceptron resembles very closely to
that obtained in numerical simulations of ellipsoids~\cite{schreck2012},
where again $D(\omega)$ consists of three well-separated regions. To
uncover further similarity, we discuss the scaling behavior of the
characteristic frequencies, $\omega_0$, $\omega_1$, and $\omega_2$.
\begin{figure}
\centering \includegraphics[width=.9\textwidth]{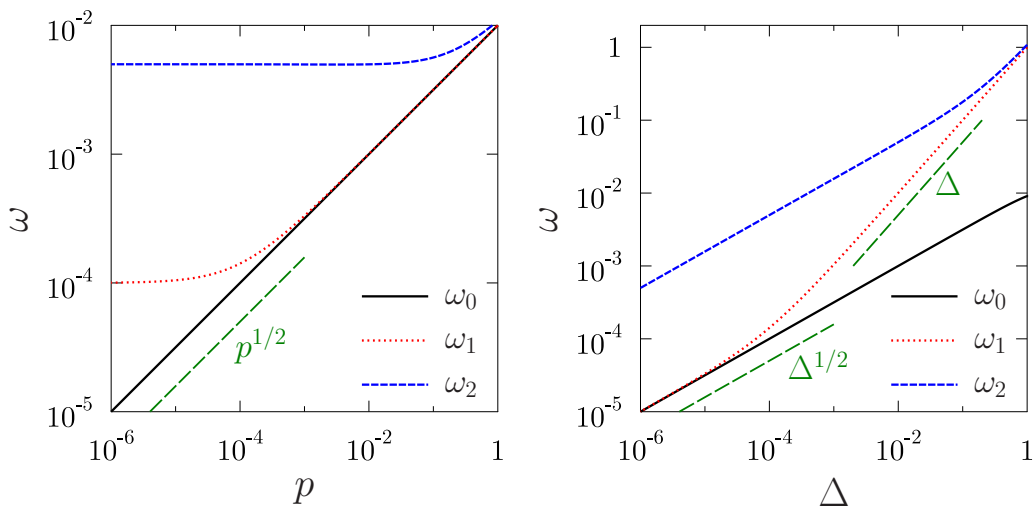}
 \caption{Scaling of the characteristic frequencies near the
 jamming transition point. The left panel shows the $p$ dependence of
 the characteristic frequencies for fixed $\Delta=10^{-4}$, while the
 right panel shows the $\Delta$ dependence for $p=10^{-4}$.
 The black solid line, red dotted line, and blue dashed lines indicate the
 theoretical results for $\omega_0$, $\omega_1$, and $\omega_2$, respectively.
 The green lines are guide for eye. }
 \label{184947_2Feb19}
\end{figure}
In Fig.~\ref{184947_2Feb19}, we show the $p$ and $\Delta$ dependence of
the characteristic frequencies, from which one can deduce the following
scaling behavior:
\begin{align}
 \omega_0&\sim \Delta^{1/2}p^{1/2},&
 \omega_1 &\sim
\begin{cases}
 \Delta^{1/2}p^{1/2} & (p\gg \Delta)\\
 \Delta & (p\ll \Delta)
\end{cases},&
\omega_2 &\sim \Delta^{1/2}.\label{193543_2Feb19}
\end{align}
The same scaling can also be obtained directly by the asymptotic
analysis of $\lambda_1,\cdots,\lambda_4$. For $p\ll\Delta$ we find the
same scaling reported in Ref.~\cite{schreck2012} for the case of
ellipsoids.  Instead it seems that the aspect ratios used in
Ref.~\cite{schreck2012} are too large to observe the scaling for $p\gg
\Delta$. It would be interesting to redo the numerical simulation to
further investigate this case.  For the lowest frequency regime, we get
that $\rho(\lambda)\sim \sqrt{\lambda}\theta(\lambda)$, which leads to
$D(\omega)\sim \omega^2$, see Fig.~\ref{185025_31Jan19} as in the case
of the spherical perceptron \cite{franz2015}.

\section{Summary and discussions}
\label{155216_12Feb19}

\begin{figure}
\centering \includegraphics[width=.8\textwidth]{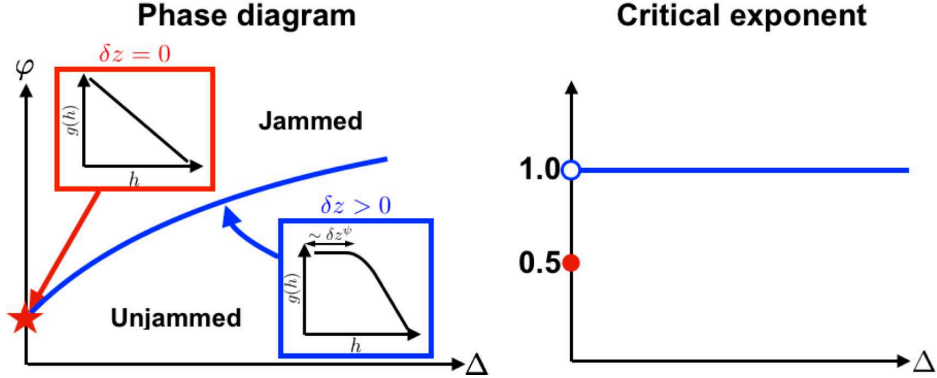} \caption{
Summary of theoretical results.  (left) The putative jamming line as a
function of the deviation from sphere $\Delta$. The red star at
$\Delta=0$ indicates that the jamming transition point of spherical
particles, where the system is isostatic $\delta z=0$ and the gap
distribution shows the power law behavior. The blue line for $\Delta>0$
represents the jamming transition point of nonspherical particles, where
the system is not isostatic $\delta z>0$ and the power law of the gap
distribution is truncated at $h\sim \delta z^\psi$.  (right) The critical
exponent of the shear modulus and the contact number near the jamming
transition point. The critical exponent jumps at $\Delta=0$ from $0.5$
to $1.0$. } \label{175547_18Feb19}
\end{figure}

In this manuscript, we constructed a mean field theory for the jamming
transition of nonspherical particles based on the analytical solution of
the polydisperse spherical perceptron model. This model is an extension
of the spherical perceptron model introduced in \cite{franz2016simplest}
to study jamming of spherical particles. In order to take into account
the particles shape, which results in internal degrees of freedom such
as the orientations, we introduced additional internal degrees of
freedom in the perceptron model~\cite{brito2018universality}. We can
parametrize the asphericity of particles using an additional control
parameter $\Delta$. We solved the model through the replica method and
we determined the phase diagram and the asymptotic behavior of the gap
(and force) distributions near and at the jamming transition point.  The
resulting generic picture is showed in the left panel of
Fig.~\ref{175547_18Feb19}, where we summarize the phase diagram
predicted by our theoretical calculation.  For $\Delta=0$, which
corresponds to spherical particles, the system is isostatic at the
jamming point, and the gap and force distributions show a power law
behavior \cite{franz2016simplest}. On the contrary, for $\Delta>0$,
which corresponds to nonspherical particles of finite asphericity, the
system is not isostatic at the jamming transition point, and the gap and
force distributions remain regular and finite. In the right panel of
Fig.~\ref{175547_18Feb19}, we show the critical exponents of the shear
modulus and contact number. Due to the regularity of the distribution
function, the critical exponent takes a rather simple value for
$\Delta>0$: the shear modulus and contact number are linearly
proportional to the proximity to the jamming transition point. On the
contrary, they are proportional to the square root of the proximity to
the jamming transition point for spherical particles. Thus, the exponent
jumps from one-half to one at $\Delta=0$. We also calculated the density
of states $D(\omega)$ and found that our model reproduces the scaling
behavior of $D(\omega)$ of ellipsoids near the jamming transition point,
previously reported in numerical simulations. Our results highlight the
specificity of the universality of jamming of spherical particles, which
holds only when the asphericity is precisely zero, whereas the
universality of nonspherical particles is far more general as it holds
no matter how small the asphericity is.

There are still several important points that deserve further
investigation. Here we give a tentative list:
\begin{itemize}
 \item In previous research, we confirmed our theoretical prediction
for the gap distribution function by computer simulation of
breathing particles, which belong to the same universality class of
       nonspherical particles~\cite{brito2018universality}. It would be
       important to repeat the same analysis for various shapes of
       nonspherical
particles to confirm the absence of criticality also in those cases.

 \item $D(\omega)$ of nonspherical particles differs dramatically from
       that of spherical particles.  An interesting question is how this
       difference affects the steady-state rheology near the jamming transition
       point. For spherical particles, it has been shown that the minimal
       eigenfrequency of $D(\omega)$ controls the viscosity near the jamming
       transition point~\cite{lerner2012unified}. It would be interesting to
       extend the theory in Ref.~\cite{lerner2012unified} to the present
       case to study how the rheology of nonspherical particles differs
       from that of spherical ones.

 \item For the low frequency regime of the density of states, our mean
       field model predicts a quadratic scaling $D(\omega)\sim
       \omega^2$.  However, in finite dimensions, it has been shown
       theoretically that spatial heterogeneity modifies the
       mean-field result below some characteristic frequency, and the
       quadratic scaling is replaced by a quartic scaling
       $D(\omega)\sim\omega^4$~\cite{ikeda2019}. This is indeed
       consistent with numerical observations in breathing
       particles, which can be considered as nonspherical particles of
       small asphericity~\cite{kapteijns2019}. It would be interesting
       future work to test if the same quartic scaling holds for
       ellipsoids and other nonspherical models.

 \item In this manuscript, we investigated the
fully connected mean-field model, which does not have a spatial
structure.  A natural next step is to include spatial fluctuations and
study the divergence of the correlation length near the jamming transition
point. This is possible by considering a Kac version of the model, as previously
done for the $p$-spin spherical model~\cite{franz2007analytic} and hard
spheres~\cite{ikeda2015one}. 

 \item Although the scaling behavior of our model near the jamming
transition point is entirely different from that of the original
perceptron model, the two models have qualitatively the same phase
diagram, in particular, the jamming line is surrounded by the replica
symmetric breaking (RSB) line for $\sigma<0$, where the model can be
mapped into a particle system. This means that RSB occurs before
 the jamming transition point is reached. It would be interesting to see if
 signatures of such a transition are present for models of nonspherical particles
 in finite dimension.

 \item Our approach is justified only for small enough asphericity. For
larger asphericity, completely new effects may appear such as (partial)
orientational ordering, orientational jamming vs. positional jamming,
additional many-body correlations, higher order corrections, those of
which can not be captured by our model.  To discuss those effects, it is
tempting to extend the present calculation for the perceptron to more
realistic models of nonspherical particles. A possibility in this
direction is to solve nonspherical particle models in the large
dimensional limit where one can write down the analytical expression of
the free energy as a functional of the density of particle positions and
angles, as shown in Ref.~\cite{yoshino2018translational}.
\end{itemize}

\ack This paper is dedicated to Giorgio Parisi in his 70th birthday;
Giorgio is a great source of inspiration for all of us and discussions
with him, as well as previous joint work, have been crucial for the
development of this research.  We particularly thank C.~Brito and
M.~Wyart for previous joint research~\cite{brito2018universality} on
which this work is based.  We also thank B.~Chakraborty, A.~Ikeda,
J.~Kurchan, S.~Nagel, S.~Franz, and H.~Yoshino for interesting
discussions related to this work.

This project has received funding from the European Research Council
(ERC) under the European Union's Horizon 2020 research and innovation
programme (grant agreement n.723955-GlassUniversality).  This work was
supported by ``Investissements
d'Avenir'' LabEx PALM (ANR-10-LABX-0039-PALM)(P. Urbani).

\appendix

\section{Full RSB free energy and saddle point equations}
\label{124714_14Feb19}
Here, we briefly explain the derivation of the replicated free-energy
and saddle point equations. For a more complete discussion see Ref.~\cite{franz2017}.

Substituting a hierarchical replica ansatz \cite{mezard1987} into the
free energy of Eq.~(\ref{113413_14Jan19}), we obtain
\begin{equation}
\begin{split}
 -\beta f &= \lim_{n\to 0}\frac{\log Z^n}{nN}
 = \frac{1}{2}\left[
 \log(1-q_M) + \frac{q_m}{\lambda(0)} + \int_0^1 dx \frac{\dot{q}(x)}{\lambda(x)}
 \right]
 \\
 &-\frac{\alpha}{2}\log\tk
 + \alpha \left.\gamma_{q_m}*f(0,h)\right|_{h=-\sigma},
\end{split}
\end{equation}
where the dot denotes derivation with respect to $x$, and
\begin{align}
 \lambda(x) &= 1- x q(x) -\int_x^1 dy q(y).
\end{align}
The function $f(x,h)$ is obtained as the solution of the Parisi equation \cite{Pa80}:
\begin{align}
 \dot{f}(x,h) &= -\frac{1}{2}\dot{q}(x)\left[f''(x,h) + x f'(x,h)^2\right],\ x_m<x<x_M,\label{134646_2Feb19}
\end{align}
with the boundary condition:
\begin{align}\label{eqf1hApp}
 f(1,h) &= \log_{1-q_M+\Delta^2/\tk}*e^{-\beta v(h)}.
\end{align}
We assume that for $x\not\in [x_m,x_M]$, $\dot{q}(x)=0$ while $q(x)$ takes a constant
value, $q(x) = q_m$ for $x\in[0,x_m)$ and $q(x)=q_M$ for $x\in (x_M,1]$.
For $x\in [x_m,x_M]$, $q(x)$ is a monotonic function, and thus one can
define the inverse function $x(q)$. Using $x(q)$,
Eq.~(\ref{134646_2Feb19}) reduces to
\begin{align}
 \dot{f}(q,h) = -\frac{1}{2}\left[f''(q,h)+ x(q)f'(q,h)^2\right],\label{140029_2Feb19}
\end{align}
with
\begin{align}
 f(q_M,h) = \log_{1-q_M + \Delta^2/\tk}*e^{-\beta v(h)}.\label{140035_2Feb19}
\end{align}
The replicated free energy is
{\medmuskip=0mu
\thinmuskip=0mu
\thickmuskip=0mu
\begin{align}
-\beta f[x(q)] &= \frac{1}{2}\left[\log(1-q_M) + \frac{q_m}{\lambda(q_m)}
 + \int_{q_m}^{q_M}\frac{dq}{\lambda(q)}\right]
 -\frac{\alpha}{2}\log \tk + \alpha \left.\gamma_{q_m}*f(q_m,h)\right|_{h=-\sigma},\label{135858_2Feb19}
\end{align}}
where
\begin{align}
 \lambda(q) &= 1-q_M + \int_q^{q_M}dp x(p).
\end{align}
In order to compute the saddle point for $q(x)$ or equivalently for $x(q)$, one should extremize
Eq.~(\ref{135858_2Feb19}) w.r.t $x(q)$, with the constraint that
$f(q,h)$ satisfies Eq.~(\ref{140029_2Feb19}) and (\ref{140035_2Feb19}).
To do this one can introduce the Lagrange multiplier $P(q,h)$  \cite{sommers1984distribution} as, 
\begin{align}
 -\beta f[x(q)] &= \frac{1}{2}\left[\log(1-q_M)+\frac{q_m}{\lambda(q_m)}+ \int_{q_m}^{q_M}\frac{dq}{\lambda(q)}\right]
 -\frac{\alpha}{2}\log\tk + \alpha \left.\gamma_{q_m}*f(q_m,h)\right|_{h=-\sigma}\new
 &-\alpha \int dh P(q_M,h)\left[f(q_M,h)-\log\gamma_{1-q_M+\Delta^2/\tk}*e^{-\beta v(h)}\right]\new
 &+\alpha \int dh \int_{q_m}^{q_M}dqP(q,h)\left\{\dot{f}(q,h)+\frac{1}{2}\left[f''(q,h)+x(q)f'(q,h)^2\right]\right\}.\label{090400_4Feb19}
\end{align}
Note that the saddle point conditions for $P(q,h)$ and $P(q_M,h)$
correctly reproduce Eq.~(\ref{140035_2Feb19}) and (\ref{eqf1hApp}),
respectively. The equations for the Lagrange multiplier $P(q,h)$ is
obtained taking the functional derivatives w.r.t. $f(q,h)$ and $f(q_m,h)$:
\begin{align}
 P(q_m,h) &= \gamma_{q_m}(h+\sigma),\new
 \dot{P}(q_M,h) &= \frac{1}{2}\left[P''(q,h)-2x(q)\left(P(q,h)f'(q,h)\right)'\right],\ q_m < q< q_M.
\end{align}
The saddle point condition for $x(q)$ is
\begin{align}
 \frac{q_m}{\lambda(q_m)^2} + \int_{q_m}^q \frac{dp}{\lambda(p)^2} = \alpha \int dh P(q,h)f'(q,h)^2.\label{141542_2Feb19}
\end{align}
In the continuous RSB phase, $x(q)$ is a continuous function for $x\in
[q_m,q_M]$, which allows us to calculate derivation of Eq.~(\ref{141542_2Feb19}) w.r.t $q$.
The first and second derivatives lead to
\begin{align}
 \frac{1}{\lambda(q)^2} &= \alpha \int dh P(q,h)f''(q,h)^2,\label{091849_4Feb19}\\
 x(q) &= \frac{\lambda(q)}{2}\frac{\int dhP(q,h)f'''(q,h)^2}{\int dh P(q,h)\left[f''(q,h)^2 + \lambda(q)f''(q,h)^3\right]}.
\end{align}
Also, the spherical constraint $\sum_{\mu=1}^MR_\mu^2 =
M$, gives the last saddle point equation for $\tilde k$
\begin{align}
 1 &= \frac{1}{\tk} +
 \frac{\Delta^2}{\tk^2}\int dh P(q_M,h)\left[f''(q_M,h) + \left(f'(q_M,h)\right)^2\right].\label{091857_4Feb19}
\end{align}

\section{Replica symmetric analysis in the unjammed phase}
\label{173919_15Feb19}
Here we investigate the replica symmetric (RS) saddle point equations in
the unjammed phase.  In this case one has that
$e^{-\beta v(h)}= \theta(h)$ so that
\begin{align}
 f_{\rm RS}(q,h) = \log \gamma_{1-q+\Delta^2/\tk}\theta(h)
 = \log\Theta\left(\frac{h}{\sqrt{2(1-q+\Delta^2/\tk)}}\right),
\end{align}
where $\Theta(x)=(\erf(x)+1)/2$.  Approaching the jamming point one has that
$q\to 1$ and $\tk\to\infty$.  In this limit, using the asymptotic
expansion of the error function, we can show that
\begin{align}
 f_{RS}(q,h) \sim -\frac{h^2}{2(1-q+\Delta^2/\tk)}\theta(-h).
\end{align}
Substituting this expression into Eqs.~(\ref{020701_17Jan19}) and 
Eq.~(\ref{091857_4Feb19}),
we obtain
\begin{align}
\tk &\sim \frac{\Delta}{(1-q)\sqrt{\alpha}}.
\end{align}
and
\begin{align}
\alpha_J = \left(\frac{1}{\sqrt{G(\sigma)}-\Delta}\right)^2.
\end{align}
which coincides with Eq.~(\ref{161700_15Jan19}) obtained from the analysis in the jammed phase.

\section{Scaling in the unjammed phase}
\label{181017_15Feb19}
Here we discuss the scaling behavior when approaching to the jamming
transition point from the unjammed phase.

  \subsection{Failure of the critical solution}
We first show that the critical jamming solution for the spherical
perceptron does not work for the nonspherical perceptron model when
$\Delta>0$.  For this purpose, we introduce the following scaling
variables:
\begin{align}
 y(q) &= \varepsilon^{-1}x(q),\new
 \hf(q,h) &= \varepsilon f(q,h),\new
 \hpi(q) &= \varepsilon^{-1}\left(\lambda(q) + \frac{\Delta^2}{\tk}\right),
\end{align}
where $\varepsilon$ is the linear distance from the jamming transition
point.  In the critical solution~\cite{franz2017}, one assumes
{\medmuskip=-1mu
\thinmuskip=-1mu
\thickmuskip=-1mu
\begin{align}
 1-q_M &\sim \varepsilon^{\kappa},& \frac{\tk}{\Delta^2} &\sim \varepsilon^{-\kappa},
\end{align}
\begin{align}
 y(q) &\sim y_J (1-q+\Delta^2/\tk)^{-1/\kappa},\new
 P(q,h) &\sim
 \begin{cases}
  (1-q+\Delta^2/\tk)^{(1-\kappa)/\kappa}p_{-}\left[h(1-q+\Delta^2/\tk)^{(1-\kappa)/\kappa}\right] & h\sim -(1-q+\Delta^2/\tk)^{(\kappa-1)/\kappa},\\
(1-q+\Delta^2/\tk)^{-a/\kappa}p_0\left(\frac{h}{\sqrt{1-q+\Delta^2/\tk}}\right) &  \abs{h} \sim \sqrt{1-q+\Delta^2/\tk}\\
p_{+}(h) & h\gg \sqrt{1-q+\Delta^2/\tk},
 \end{cases}
\end{align}}
and
\begin{align}
 m(q,h) &= \hpi(q)\hf'(q,h) = -\sqrt{1-q+\Delta^2/\tk}\M\left(\frac{h}{\sqrt{1-q+\Delta^2/\tk}}\right),\new
 \M(t\to\infty) &=0,\ \M(t\to-\infty) =t,
\end{align}
where
\begin{align}
 \kappa &= 1.41574,& a &= 1-\frac{\kappa}{2}.\label{123635_5Feb19}
\end{align}
Substituting these relations into Eq.~(\ref{091857_4Feb19}), we get 
\begin{align}
 1 &\sim \frac{\Delta^2}{\tk^2}\int dh P(1,h)f'(1,h)^2 = \frac{\Delta^2}{\tk^2\varepsilon^2\hpi(1)^2}\int dh P(1,h)m(1,h)^2,
\end{align}
which leads to 
\begin{align}
\Delta^2 \sim \varepsilon^{2\kappa-2}\int_{-\infty}^0 dt p_{-}(t)t^2.
\end{align}
This equation can be satisfied only when $\Delta=0$, i.e., in the case
of the standard perceptron. A similar equation can be also obtained on
approaching to the jamming transition from the jammed phase.
We deduce that the critical jamming solution is inconsistent when $\Delta>0$.

\subsection{Scaling solution for $\Delta>0$}
The jamming scaling for $\Delta>0$ is describe by the regular full RSB
solution. We define the scaling solutions as
\begin{align}
 1-q_M &= \varepsilon,\new
 \tk &= \frac{\Delta^2}{\varepsilon}K.\label{142742_5Feb19}
\end{align}
For $\varepsilon\ll 1$, we have
\begin{align}
 f(1,h) \sim -\frac{h^2}{\varepsilon(1+1/K)}\theta(-h).
\end{align}
Also we assume that $P(q,h)$ is a regular and finite function.  Then,
Eq.~(\ref{091857_4Feb19}) reduces to
\begin{align}
 1 = \frac{\Delta^2}{\tk^2}\int dh P(1,h)f'(1,h)^2 \sim \frac{1}{\Delta^2(1+K)^2}\int_{-\infty}^0 dh P(1,h)h^2.
\end{align}
Thus, the regular scaling solution gives a well-defined expression in
the $\varepsilon\to 0$ limit, unlike the critical scaling.  Similarly,
the pressure $p$ can be obtained from
\begin{align}
 p &= \frac{1}{\alpha}\diff{f}{\sigma } = T\int dh P(1,h)f'(1,h) \sim \frac{T}{\varepsilon(1+1/K)}\int_{-\infty}^0 dh P(1,h)h,
\end{align}
which implies that $p\sim \varepsilon^{-1}$.  Combining this with
Eqs.~(\ref{142742_5Feb19}), we can determine the pressure dependence of
$q_M$ and $\tk$ for $p\ll 1$:
\begin{align}
 1-q_M &\sim p^{-1},& \tk &\sim p.
\end{align}

\section{Derivation of the eigenvalue distribution}
\label{160922_16Feb19}

We want to compute the eigenvalue distribution $\rho(\lambda)$ of the
$(N+zN)\times (N+zN)$ dimensional Hessian defined by:
\begin{align}
 \he  &=
 \begin{pmatrix}
  H & T\\
  T^t & Q
 \end{pmatrix},
\end{align}
where 
\begin{align}
H_{ij} &\equiv \pdiff{^2 V_N}{x_i\partial x_j} = \frac{1}{N}\sum_{\mu=1}^{zN}\xi_i^\mu\xi_j^\mu-\zeta\delta_{ij},\new
Q_{\mu\nu}  & \equiv \pdiff{^2 V_N}{R_\mu\partial R_\nu} = \left[\Delta^2+k_R\right]\delta_{\mu\nu},\new
 T_{\mu i} & \equiv \pdiff{^2 V_N}{R_\mu\partial x_i } =-\frac{\Delta \xi_i^\mu}{\sqrt{N}},
\end{align}
for $i,j=1,\cdots,N$ and $\mu,\nu=1,\cdots, zN$. 
We perform the computation using the
Edwards-Jones formula~\cite{edwards1976}:
\begin{align}
 \rho(\lambda) &\equiv \frac{1}{N+zN}\sum_{k=1}^{N+zN}
 \delta(\lambda-\lambda_k)
 = -\frac{2}{(N+zN)\pi}\lim_{\varepsilon\to 0}
 \im\pdiff{}{\lambda}\overline{\log Z(\lambda-i\varepsilon)},
\end{align}
where the overline denotes the average over the quenched disorder
$\xi_i^\mu$, and we have introduced the partition function as 
\begin{align}
 Z(\lambda) &= \int d\bu
 \exp\left[-\frac{1}{2}\bu\cdot(\lambda I_{N+zN}-\he)\cdot\bu\right].
\end{align}
Here $I_d$ denotes the $d$-dimensional identity matrix.
Performing the Gaussian integration we get 
\begin{align}
 \log Z(\lambda) &= -\frac{1}{2}\log\det\left[\lambda I-\he\right]\new
 &= -\frac{1}{2}\left\{
 \log\det(\lambda I_{zN} -Q) + \log\det(\lambda I_N -H -T^t(\lambda I_{zN}-Q)^{-1}T)
 \right\}\new
 &= -\frac{1}{2}\left\{
 Nz\log(\lambda- \Delta^2-k_R))
  + \log\det A  \right\},\label{174656_29Jan19}
\end{align}
where we have introduced a $N\times N$ matrix as 
\begin{align}
 A_{ij} &= (\lambda+\zeta)\delta_{ij} - \frac{a(\lambda)}{N}\sum_{\mu=1}^{Nz}\xi_i^\mu\xi_j^\mu,\new
 a(\lambda) &= \left(1 + \frac{\Delta^2}{\lambda-\Delta^2-k_R}\right).
\end{align}
Replacing the quenched average by the annealed one and using the saddle
point method, we can show that
\begin{align}
 \overline{\log \det A} &\approx \log\overline{\det A}\new
 &\sim  N(\lambda+\zeta)q -N \log q + Nz\log(1-a(\lambda)q),\label{174714_29Jan19}
\end{align}
where $q = N^{-1}\sum_{i=1}^N u_i^2$ is to be determined by the saddle point condition:
\begin{align}
 \lambda + \zeta -\frac{1}{q} -z\frac{a(\lambda)}{1-\alpha(\lambda)q} = 0.
\end{align}
This can be solved for $q$:
\begin{align}
 q(\lambda) &= \frac{1}{2a(\lambda)} + \frac{1-z}{2(\lambda+\zeta)} \pm
 \frac{\sqrt{\left[\lambda+\zeta -a(\lambda)(1+\sqrt{z})^2\right]\left[\lambda+\zeta-a(\lambda)(1-\sqrt{z})^2\right]}}
 {2(\lambda+\zeta)a(\lambda)},
\end{align}
where the sign is to be chosen so that $\rho(\lambda)$ is positive and
normalized. Substituting Eqs.~(\ref{174656_29Jan19}) and
(\ref{174714_29Jan19}) into Eq.~(\ref{174738_29Jan19}), we have
\begin{align}
 \rho(\lambda) &= \frac{1}{\pi}\lim_{\varepsilon\to 0}\im G(\lambda-i\varepsilon),\new
 G(\lambda)&\equiv \frac{z}{1+z}\frac{1}{\lambda-\Delta^2-k_R} + \frac{1}{1+z}q(\lambda)
 -\frac{z}{1+z}\frac{a'(\lambda)}{1-a(\lambda)q(\lambda)}q(\lambda).
\end{align}
After a straightforward calculation, we finally get
{\medmuskip=0mu
\thinmuskip=0mu
\thickmuskip=0mu
\begin{align}
 \rho(\lambda) &= \frac{1-z}{1+z}\theta(1-z)\delta(\lambda+\zeta)
 + \frac{z-1}{1+z}\theta(z-1)\delta(\lambda-\Delta^2 -k_R)\new
 &+ \abs{\frac{\im \sqrt{\prod_{i=1}^4(\lambda-\lambda_i)}}{2\pi(1+z)}
 \left[
 \frac{1}{(\lambda-k_R)(\lambda+\zeta)}
 -\frac{1}{(\lambda-k_R)^2} + \frac{1}{(\lambda-k_R)(\lambda-\Delta^2-k_R)}
 \right]},\label{170620_30Jan19}
\end{align}}
where
{\medmuskip=0mu
\thinmuskip=0mu
\thickmuskip=0mu
\begin{align}
 \lambda_1 &= \frac{1}{2}\left[
 (1-\sqrt{z})^2 + k_R + \Delta^2 -\zeta -\sqrt{\left\{(1-\sqrt{z})^2 + k_R + \Delta^2 -\zeta\right\}^2
 -4 \left\{(1-\sqrt{z})^2k_R-(k_R+\Delta^2)\zeta\right\}}
 \right],\new
 \lambda_2 &= \frac{1}{2}\left[
 (1+\sqrt{z})^2 + k_R + \Delta^2 -\zeta -\sqrt{\left\{(1+\sqrt{z})^2 + k_R + \Delta^2 -\zeta\right\}^2
 -4 \left\{(1+\sqrt{z})^2k_R-(k_R+\Delta^2)\zeta\right\}}
 \right],\new
 \lambda_3 &= \frac{1}{2}\left[
 (1-\sqrt{z})^2 + k_R + \Delta^2 -\zeta +\sqrt{\left\{(1-\sqrt{z})^2 + k_R + \Delta^2 -\zeta\right\}^2
 -4 \left\{(1-\sqrt{z})^2k_R-(k_R+\Delta^2)\zeta\right\}}
 \right],\new
  \lambda_4 &= \frac{1}{2}\left[
 (1+\sqrt{z})^2 + k_R + \Delta^2 -\zeta +\sqrt{\left\{(1+\sqrt{z})^2 + k_R + \Delta^2 -\zeta\right\}^2
 -4 \left\{(1+\sqrt{z})^2k_R-(k_R+\Delta^2)\zeta\right\}}
 \right].\label{173718_16Feb19}
\end{align}}
Since $z>1$ in the RSB jammed phase, Eq.~(\ref{170620_30Jan19}) can be
slightly simplified as
{\medmuskip=0mu
\thinmuskip=0mu
\thickmuskip=0mu
\begin{align}
 \rho(\lambda) &= 
 \frac{z-1}{1+z}\delta(\lambda-\Delta^2 -k_R)\new
 &+ \abs{\frac{\im \sqrt{\prod_{i=1}^4(\lambda-\lambda_i)}}{2\pi(1+z)}
 \left[
 \frac{1}{(\lambda-k_R)(\lambda+\zeta)}
 -\frac{1}{(\lambda-k_R)^2} + \frac{1}{(\lambda-k_R)(\lambda-\Delta^2-k_R)}
 \right]}.
\end{align}}

\section{Calculation of $\zeta$}
\label{161701_16Feb19}
We here determine the Lagrange multiplier $\zeta$ introduced in
Sec.~\ref{185025_31Jan19}.
For this purpose, we rewrite the free energy Eq.~(\ref{090400_4Feb19}) as 
\begin{align}
 -\beta f[x(q)] &= \frac{1}{2}\left[\log(q_d-q_M)+\frac{q_m}{\lambda(q_m)}+ \int_{q_m}^{q_M}\frac{dq}{\lambda(q)}\right]
 -\frac{\alpha}{2}\log\tk + \alpha \left.\gamma_{q_m}*f(q_m,h)\right|_{h=-\sigma}\new
 &-\alpha \int dh P(q_M,h)\left[f(q_M,h)-\log\gamma_{q_d-q_M+\Delta^2/\tk}*e^{-\beta v(h)}\right]\new
 &+\alpha \int dh \int_{q_m}^{q_M}dqP(q,h)\left\{\dot{f}(q,h)+\frac{1}{2}\left[f''(q,h)+x(q)f'(q,h)^2\right]\right\}\new
 &+\frac{\beta\zeta N }{2}\left(q_d-1\right),
\end{align}
where $q_d=N^{-1}\sum_{i}x_i^2$.
The saddle point condition for $q_d$ leads to
{\medmuskip=0mu
\thinmuskip=0mu
\thickmuskip=0mu
\begin{align}
 &\frac{1}{2}\left(\frac{1}{q_d-q_M}-\frac{q_m}{\lambda(q_m)^2}-\int_{q_m}^{q_M}\frac{dp}{\lambda(p)^2}\right)
 + \alpha \int dh P(q_M,h)\pdiff{}{q_d}\log\gamma_{q_d-q_M+\Delta^2/\tk}*e^{-\beta v(h)}\new
 &+ \frac{\beta \zeta}{2} =0 
\end{align}}
After some manipulations, this can be rewritten as 
\begin{align}
 \beta\zeta &= -\frac{1}{q_d-q_M} + \alpha \int dh P(q_M,h)f''(q_M,h).
\end{align}
Substituting $q_M\sim q_d-T\chi$, $f(q_M,h)\sim -\beta
h^2\theta(-h)/(1+\chi + \Delta^2/k_R)$, and taking the zero temperature
limit $T\to 0$, we get
\begin{align}
 \zeta &= \frac{1+\Delta^2/k_R}{\chi} = \frac{(\sqrt{z}-1)^2}{1+\Delta^2/k_R}.\label{173823_16Feb19}
\end{align}

 \section*{References}

\providecommand{\newblock}{}


\end{document}